\documentclass[prb, aps, superscriptaddress, onecolumn,12pt]{revtex4}

\usepackage{amsmath}
\usepackage{array}
\usepackage{graphicx}
\usepackage{amstext}
\usepackage{amsfonts}
\usepackage{amsmath,amssymb}

\usepackage{bm}
\usepackage{diagbox}

\newcommand{\be}{\begin{eqnarray}}
\newcommand{\ee}{\end{eqnarray}}
\newcommand{\ba}{\begin{array}}
\newcommand{\ea}{\end{array}}
\newcommand{\bml}{\begin{mathletters}}
\newcommand{\eml}{\end{mathletters}}
\newcommand{\la}{\langle}
\newcommand{\ra}{\rangle}
\newcommand{\lb}{\left(\begin{array}}
\newcommand{\rb}{\end{array}\right)}

\newcommand{\bmu}{{\boldsymbol \mu}}

\begin{document}

\author{Jianshu Cao}
\email{jianshu@mit.edu}
\affiliation{Department of Chemistry, Massachusetts Institute of Technology, Massachusetts, 02139 USA }

\title{Generalized resonance energy transfer theory: Applications to vibrational energy flow in optical cavities} 

\date{\today}

\begin{abstract}

A general rate theory for resonance energy transfer is formulated to incorporate any degrees of freedom (e.g., rotation, vibration, exciton, and polariton) 
 as well as  coherently-coupled composite states.  The compact rate expression allows us to establish useful relationships:
(i)  detailed balance condition when the donor and acceptor are at the same temperature;
(ii)  proportionality to the overlap between donor's emission and acceptor's absorption spectra;
(iii)  scaling with the effective coherent size, i.e., the number of coherently coupled molecules;
(iv) spatial and orientational dependences as derived from the interaction potential.
When applied to cavity-assisted vibrational energy transfer, the rate formalism provides an intuitive and quantitative explanation of
intriguing phenomena such as cooperativity, resonance, and nonlinearity in the collective vibrational strong coupling regime,
as demonstrated in recent simulations.

\end{abstract}

\maketitle

\section{Introduction}

Energy transfer is a fundamental process in chemical physics and plays a pivotal role in photosynthesis, photovoltaics, and chemical reactions in general.  
Much of our understanding is derived on the basis of a specific energy form, such as Forster resonance energy transfer (FRET) for excitons\cite{forster65,cao193} and inter-molecular or intra-molecular energy relaxation for vibrations,\cite{bigwood98,baiz20} and is mostly limited to independent donors or acceptors.  
The first goal of this paper is the formulation of a generalized resonant energy transfer (gRET) theory, which features two-fold generalizations: 
(i) applicability to any resonance quantum transfer processes including electronic, vibrational, rotational transitions as well as hybrid energy transfer processes
such as vibronic resonance and polariton resonance and (ii) extension from individual molecules to a collection of molecules
coherently coupled via direct interactions or optical cavity fields.  
The compact gRET rate expression allows us to establish its detailed balance (DB) condition, proportionality to the spectral overlap,  scalings with 
the effective coherent size, and configurational dependence.

The second goal of the paper is the application of gRET to describe  vibrational energy flow in optical cavities.  As demonstrated in recent 
experiments,\cite{george16,thomas16,thomas19,lather19,hirai20a,dunkelberger16,xiang18,xiang20} the electromagnetic field of an optical cavity can mix with vibrational modes of molecular systems to collectively form vibrational polaritons and 
thus facilitate vibrational energy transfer (VET) or even modify chemical kinetics in the vibrational strong coupling (VSC) regime.
These intriguing findings have stimulated theoretical and numerical studies,
including molecular dynamics simulations of VET in cavities.\cite{flick17a,schafer21,li20b,li21}
In particular, to develop a basic understanding of cavity-assisted VET, we apply the gRET rate expression
to quantitatively explain several observations reported in a recent cavity molecular dynamics simulation.\cite{li21}

\section{Generalized RET}

The generalized resonance energy transfer (gRET) between
a donor (D) and an acceptor (A)  as illustrated by Fig.~1(a) is described by the total Hamiltonian
\begin{equation}
H_{total} = H + H_{DA}= H_D + H_A + H_{DA},
\notag
\end{equation}
where $H_{DA}$ is the coupling between the donor
and acceptor,  $H_{A(D)}$ is the acceptor (donor) Hamiltonian,  
 $H=H_A+H_D$ is the uncoupled part of the total Hamiltonian.
 In a multi-chromophoric (MC) system, the donor or acceptor state
can represent an aggregate of coherently coupled molecules
and is therefore not limited to monomers.
The interaction Hamiltonian between the donor and acceptor is given by
\begin{equation}
	H_{DA} =   {  \boldsymbol \mu }^-_D { \bf J}  { \boldsymbol \mu }^+_A  
	= \sum^{N_A}_{m=1} \sum^{N_D}_{n=1}     {\vec \mu}^-_{D,m}  {\cal J}_{mn}  {\vec \mu}^+_{A, n} 
\label{eq:II_9}
\end{equation}
where ${\cal J}_{mn}$ depends on the intermolecular configuration and will be treated perturbatively.
For N molecules in 3 dimensional space, 
${\boldsymbol \mu } = \{  \vec{\mu}_n \}$  is the 3N dipole vector  and ${\bf J} = \{ {\cal J}_{mn} \}$ is the (3N)x(3N)  coupling tensor.
Hereafter, the bold font represents vectors and tensors in the 3N dimensional space and the calligraphic font represents tensors 
in the 3 dimensional space.

The acceptor (donor) Hamiltonian is a function of  
the molecular degrees of freedom as well as 
solvent, cavity, and all the other degrees of freedom
associated with the acceptor (donor) state. 
For.a vibrational degree of freedom, a compact notation to represent such a state is
\begin{equation}
	\left| A \right> =\left| \nu \right>_A \left| b \right>_A  \left| c \right>_A
	=  \left|  \nu_1,\dots, \nu_n, \dots \right>_A \left| b \right>_A  \left| c \right>_A
\notag
\end{equation}
where $\nu_n$ is the vibrational quantum number of the n-th molecule, $b$ represents the set of bath modes,
and $c$ represents the cavity mode(s). The donor state $\left| D \right>$ is defined similarly.

To formulate the gRET theory, we should understand
the time scales involved in the resonance energy transfer process. The rate process
describes the incoherent transfer from a donor
to an acceptor, which occurs after the donor is 
in equilibrium with its thermal bath and cavity field
or in the steady-state under constant pumping.  In other words, the donor or acceptor 
will relax to equilibrium or steady-state  on a time scale 
that is faster than the transfer time $1/k$. 
Based on the above conditions, the gRET rate can be
obtained straightforwardly from Fermi’s golden rule rate,\cite{cao143}  giving
\begin{equation}
	k = { 2\pi  \over Z_D } \sum_{initial, final} \rho_{initial}
	\left|\left<\Psi_{final} \left|H_{DA} \right|\Psi_{initial} \right>\right|^2
	\delta\left(E_{final}  - E_{initial} \right).
	\label{FGR}
\end{equation}
where the subscripts {\it initial} and {\it final} refer to the initial and final states, respectively. 
Here, $\left|\Psi\right> = \left| D \right>  \left| A \right>$ is the composite
donor-acceptor state, and $E= E_A + E_D$  is the corresponding eigen-energy associated with 
the uncoupled Hamiltonian $H=H_A+H_D$.
The initial state distribution function $\rho_{initial}$ is diagonal in the eigenstate basis
and can be factorized as $\rho_{initial}  =\rho_A\rho_D$.  $Z_D= {\rm Tr} (\rho_D)$ is the donor partition function.

To proceed, we  transform Eq. (\ref{FGR}) to the time domain and obtain
\be
	Z_D k    &=& \int^\infty_{-\infty} dt~{\rm Tr}  [ H^+_{DA}(t) H_{DA}(0)   \rho_{initial}  ]
\notag\\		
&=&	\int^\infty_{-\infty} dt~{\rm Tr}    [  \bmu^-_D(0)  \rho_D \bmu_D^+(t)]   {\bf J}  {\rm Tr} [\bmu^-_A(t)  \bmu^+_A(0) \rho_A ] {\bf J}^T   
\notag
\ee
where the trace is taken over all degrees of freedom and the time-dependence in operators arises from the interaction picture representation,
${\bmu}(t) = \exp(iHt) \bmu \exp(-iHt)$ and $H_{DA}(t) = \exp(iHt) H_{DA} \exp(-iHt)$.
To simplify the rate expression, we introduce dipole correlation tensors as
\be
	{\bf C}^>(t)	 &=&	  	{\rm Tr} [ \bmu^-(t) \bmu^+(0) \rho ] 
\notag\\
	{\bf C}^<(t)	&=&	 	{\rm Tr} [\bmu^+(0) \bmu^-(t) \rho]^T	= 	{\rm Tr} [ \bmu^-(t) \rho  \bmu^+(0) ]
\notag
\ee
and the corresponding Fourier transform as  $ {\bf C}(\omega) = \int^\infty_{-\infty}dt e^{i\omega t}{\bf C}(t) $.
 These correlation tensors, ${\bf C} = \{ {\cal C}_{nn'} \}$,   
have two sets of indices, one for molecules and the other for spatial coordinates, so their dimensions are $(3N)^2$.
Then,  the gRET rate is expressed in a compact form as
\begin{equation}
	k = {1 \over Z_D}  \int^\infty_{-\infty} dt\text{ Tr } [{\bf J}^T{\bf C}^<_D(-t)\textbf{J} {\bf C}^>_A(t)] 		
	=\frac1{2\pi Z_D } \int^\infty_{-\infty} d\omega\text{ Tr}	[{\bf J}^T {\bf C}^<_D(\omega){\bf J C}^>_A(\omega)] 
\label{rate}
\end{equation}
where  $k$ is determined by the donor-acceptor
coupling {\bf J}, and the overlap integral of the acceptor's
correlation tensor  ${\bf C}^>_A(\omega)$
and the donor's correlation tensor ${\bf C}^<_D(\omega)$. 
These dipole correlation tensors are intimately related to the electronic or vibrational spectrum,\cite{mukamel99}
thus supporting the notation of spectral overlap. 
The influences of the system-bath coupling or the system-cavity coupling on the
transfer rate are reflected in the line widths and peak frequencies of these correlation tensors, 
which will be calculated later.

In general, when the system is in thermal equilibrium, i.e., $\rho= e^{-\beta H}$, 
the correlation function obeys the detailed balance (DB) relationship, $C^>(\omega)=\exp(\beta\omega\hbar) C^<(\omega)$.
Using the relation in Eq.~(\ref{rate}), we find
\be
Z_D k_{D\rightarrow A} = Z_A k_{A \rightarrow D}  
\label{DB}
\ee
which is exactly the DB condition for gRET.
The DB condition can be broken if the donor and acceptor have different temperatures and/or chemical potentials. 
Then,  the net energy flow from the donor to acceptor  defines the non-equilibrium steady-state
flux (or current),  $F = Z_D k_{D\rightarrow A} - Z_A k_{A \rightarrow D} $, 
which has been used previously in light-harvesting energy transfer\cite{cao199} as well as in various 
non-equilibrium transport settings.

We now examine the effects of quantum coherence in the donor or acceptor state
on the gRET rate.  First,  for monomers, the gRET rate in Eq.~(\ref{rate}) reduces to the RET rate,
\begin{equation}
	k_{mon} = \frac{ \rm Tr ({\cal J}^2)}{2\pi} 
		\int^\infty_{-\infty} d\omega C^<_D(\omega) C^>_A(\omega) ,
\label{monomer}
\end{equation}
where $C(t)$ is the monomer dipole correlation function and is assumed isotropic, i.e., ${\cal C}_{nn'} =  C  \delta_{nn'} {\cal I}$.  
Next, for far-field transfer,  the donor-acceptor
coupling can be taken independent of the molecular index, i.e., ${\cal \bf J}_{mn} = {\cal J} $. 
Then, the gRET rate formula remains the same as  Eq.~(\ref{monomer})
but the spectral correlation functions are defined as
\be
	C(\omega) =	\sum_{m,n} [C]_{mn}(\omega). 
\label{collective}	
\ee
Let's consider N-scalings of identical molecules without frequency resolution (i.e. ignoring the spectral overlap).
 (i) If there is no spatial coherence, then the correlation tensor is diagonal ${\bf C} = C_{mon} \delta_{mn}$
such that  $C =N C_{mon}$ and the incoherent rate is  $k_{inc}=N_A k_{mon}$. 
Here, the subscript $mon$ refers to the monomer result. 
(ii)  If the N molecules are fully coherent, then the correlation tensor is constant, $C_{mn}=C_{mon}$, 
such that $C  =N^2 C_{mon}$ and  $k= N_A^2 N_D k_{mon}= N_A N_D k_{inc}$.  
(iii) If the N molecules are partially coherent with coherent size of $N_{coh}$, then we have $C  = N N_{coh} C_{mon}$ and 
$
k =N_{A, coh} N_{D, coh} k_{inc},
$
 which applies for $N>N_{coh}$.   (iv) In reality, N should be interpreted as the number of coherently coupled molecules,
which is limited by dynamic and static disorder in cavity polaritons.\cite{cao192,cao_pre5}
 Taking the example of the exponential correlation, i.e., $C_{mn}= C_{mon} \exp(-|m-n|/N_{coh})$,
 we can define the effective coherent length
 \be
N_{eff}  =  {1 \over N}  \sum_{n,m}^N e^{-{|m-n| \over N_{coh} }} \approx  {1\over N} [ (N-N_{coh}) N_{coh} + N_{coh}^2 e^{-N/N_{coh}} ]
\label{eff}
 \ee
 which recovers the large aggregate limit of $ N_{eff} = N_{coh}$ and the small aggregate limit of $N_{eff}=N$.  
 Then, the transfer rate  is given as
 $k=  {N}_{A.eff} N_{D,eff} k_{inc}$, which interpolates between the coherent and incoherent limits.

Eq.~(\ref{rate}) is general and reduces to the FRET theory in exciton systems as a special case.
Specifically, with the electronic transition dipole $ \mu \propto  (|e \ra \la g| + |g \ra \la e|)$, 
we can identify $C^+(\omega) = I(\omega)$ as the electronic absorption spectrum,
and $C^-(\omega) = E(\omega)$ as the electronic emission spectrum. Then,  Eq~(\ref{rate}) 
becomes
\be
k_{FRET} =\frac1{2\pi} \int^\infty_{-\infty} d\omega\text{ Tr}	[{\bf J}^T {\bf E}_D(\omega){\bf J I}_A(\omega)] 
\label{FRET}
\ee 
which is exactly the generalized-FRET rate for MC systems.\cite{sumi99,scholes00,hu02,jang04,cao143}
With FRET as an example, we emphasize that 
the gRET rate formalism presented in this paper is completely general and is applicable to any resonance quantum transfer processes,
 including electronic, vibrational, and rotational transitions as well as hybrid transitions via rovibrational resonance, vibronic resonance, polariton resonance, 
 or vibrational-polariton resonance.    

  \begin{figure}
\centerline{\scalebox{0.6}{\includegraphics[trim=0 0 0 0,clip]{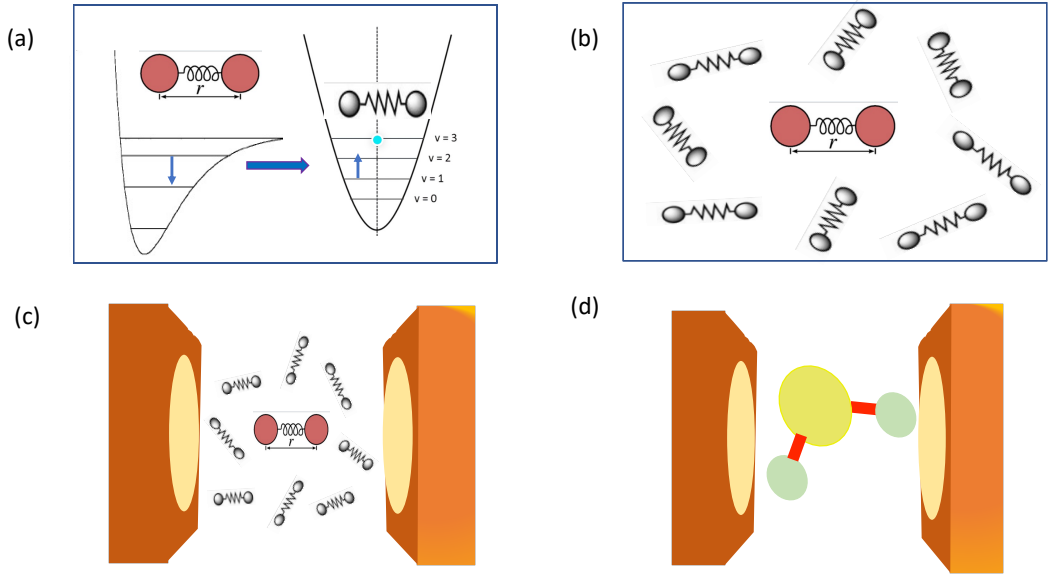}}}
\caption{Illustration of various scenarios of resonance vibrational energy transfer (VET):
(A) VET from a donor to an acceptor, both modeled as the Morse oscillator; (b) vibrational relaxation due to VET between a central molecule
and a distribution of solvent molecules; (c) cavity-assisted VET from a donor molecule to a collection of acceptor molecules coherently coupled 
by the cavity field; (d) cavity-assisted intra-molecular energy relaxation resulting from non-linear coupling between molecular modes in 
an optical cavity.}
\end{figure}

\section{Intermolecular vibrational coupling}

We consider the interaction potential between a pair of molecules, 
$U(\vec{R}_1-\vec{R}_2) = U (\vec{R}-\vec{r}_1 + \vec{r}_2) $, where $\vec{R}$ is the 
vector connecting the centers of mass of  the two molecules, and $\vec{r}_1$ and $\vec{r}_2$  are the molecular 
vibrational coordinates.  Here, the interaction is isotropic so that the potential is central symmetric.  Typically, the vibrational amplitude is 
considerably smaller than the intermolecular distance, such that  the Taylor expansion yields
\be
 U(\vec{R}_1-\vec{R}_2)= U(R) - { dU \over dR}  \hat{R} (\vec{r}_1- \vec{r}_2) + { 1 \over 2} (\vec{r}_1- \vec{r}_2)  {\cal V}_{12} (\vec{r}_1- \vec{r}_2)
 + \cdots
 \label{expansion}
\ee
where $\hat{R} = \vec{R}/R$ is the unit vector associated with $\vec{R}$.  
The coupling tensor is defined as
\be
{\cal V} = R  { d \over dR}  ( {1\over R}  { dU \over dR} ) \hat{R}  \cdot \hat{R}  +   {1\over R}  { dU \over dR} {\cal I }
\label{tensor}
\notag
\ee
where $\cal I$  is the identity tensor.  For the special case of  charge interactions,  $U(R) \propto 1/R$,  the coupling tensor reduces to the 
standard  dipole tensor, i.e.,
$
{\cal V} =  ( 3 \hat{R}  \hat{R}  - {\cal I} ) / R^3. 
$
In Eq.~(\ref{expansion}),  the first-order expansion term does not involve VET and will not be further considered. 
The second-order expansion term in Eq.~(\ref{expansion}) is rewritten as
\be
-\vec{r}_1 {\cal V}_{12} \vec{r}_2 + { 1 \over 2} (\vec{r}_1 { \cal V}_{12} \vec{r}_1 + \vec{r}_2 { \cal V}_{12} \vec{r}_2 )   
\notag
 \ee
where the first term describes resonance VET and will be the focus of discussion thereafter.
The second term represents a constant contribution of the intermolecular interaction to the site energy 
and will not be further considered. 
For the compatibility with the gRET formalism, we transform the vibrational coordinate $\vec{r}$ 
to the linear transition dipole moment $\mu$ via $\vec{\mu}= \mu' \vec{q}$ where $\mu'$ is the linear coefficient.
Then, the donor-acceptor coupling becomes
\be
(H_{DA})_{12}=\vec{\mu}_1 {\cal J}_{12}  \vec{\mu}_2 = -\vec{r}_1 {\cal V}_{12} \vec{r}_2
\ee 
where ${\cal J}_{12} = -\mu'_1 \mu'_2 {\cal V}_{12} $ is the effective dipole coupling tensor.

As a side note, the vibrational mode is generally a non-linear operator for an anharmonic potential. 
Taking the example of the Morse potential, we have 
\be
q = \sqrt{ { \hbar  \over 2 m \omega} } [(a+ a^+) - \chi (a+a^+)^2+ \cdots]
\notag
\ee
where $\chi$ denotes the anharmonicity.\cite{cao54} 
As a result, the linear excitation can induce not only single-phonon transitions but also multi-phonon transitions and dephasing.
The latter is also known as anharmonicity-induced dephasing [see Appendix A of Ref.~(\onlinecite{cao78})].  
These non-linear effects are usually small but can be amplified by the cavity-induced cooperativity.

\section{VET rate}

We apply the gRET theory  to intermolecular vibrational energy transfer  (VET) as illustrated for various cases in Fig.~1. 
For monomers, the gRET rate formula  Eq.~(\ref{rate})  simplifies to Eq.~(\ref{monomer}).  
As an example, we consider  the harmonic oscillator potential with frequency $\omega$ and dipole operator $\mu=\sqrt{\alpha} (a+a^+)$, 
where $a^-$ and $a^+$ are the lowering and raising operators respectively and $\alpha=  (\mu')^2 \hbar / (2 m  \omega) $ is the coefficient. 
Assuming slow bath relaxation, we have the Gaussian line-shape 
\be
C^>_A(\omega) &=& \alpha_A (\nu_A+1)  G(\omega-\omega_A, \lambda_A) 
\notag\\
C^<_D(\omega) &=& \alpha_D \nu_D  G(\omega-\omega_D, \lambda_D)  
\notag
\ee 
where $\lambda_A$ ($\lambda_D$) is the reorganization energy, 
$\nu_A$ ($\nu_D$) is the vibrational quantum number of the initial state, and $\omega_A$ ($\omega_B$) is the vibrational 
frequency.  Here, the Gaussian function is defined as 
\be
G(\omega, \lambda) = { 1 \over  \sqrt{ 4\pi \lambda k_B T} }  \exp(-{ \omega^2 \over 4 \lambda k_B T}) 
\notag
\ee
where $k_B T$ is the thermal energy. Then, the VET rate is given by 
\be
k = { {\rm Tr} {\cal J}^2  \over 2 \pi} \alpha_A \alpha_D \nu_D (\nu_A+1)  G(\omega_A-\omega_D, \lambda_A +\lambda_D)
\label{VET}
\ee
which exhibits the maximal transfer rate at the resonance $\omega_A=\omega_D$.   The above expressions can be easily extended
to the thermal equilibrium by introducing the thermal average, $ \la \nu \ra  = [ \exp(\beta \hbar\omega) -1 ]^{-1}$, which ensures 
the DB relation in Eq.~(\ref{DB}).  As illustrated in Fig.~1(b),  the VET rate in Eq.~(\ref{VET}) can also describe vibrational energy relaxation
by summing over all acceptors, which results in an ensemble average of molecular configurations. Specifically,
the relaxation rate can be obtained by replacing the prefactor in  Eq.~(\ref{VET}) with a configurational average:
\be
 {\rm Tr} {\cal J}^2   \rightarrow \la {\rm Tr} {\cal J}^2 \ra = \int \rho(\vec{R}) [{\rm Tr} {\cal J}^2(\vec{R})] d\vec{R} 
 \notag
 \ee
  where $\rho(\vec{R}) $ is the density distribution of acceptor molecules. More generally, we shall also account for orientational distribution in conjunction 
  with the spatial distribution.

To go beyond the monomer case, we now explicitly evaluate Eq.~(\ref{rate})  under the diagonal approximation, previously adopted to approximately
evaluate FRET rates in purple bacteria $LH_2$.\cite{hu02,cao132}  Specifically,
we consider the transfer rate between a pair of delocalized eigenstates, ignoring the entanglement between different eigenstates.  
For simplicity, we limit our discussion to far-field and explicitly evaluate the collective 
dipole correlation function defined in Eq.~(\ref{collective}) for the j-th eigenstate, giving
\be
C^>_j(\omega) =   S_j \alpha_j (\nu_j+1) G(\omega-\omega_j, \lambda_j)
\notag
\ee
and a similar expression for $C_j^<(\omega)$.
Here $S_j = \sum_{n,m} c_{n,j} c^*_{m,j}$  is a measure of the coherent size, i.e., $S \approx N_{eff}$,  and  $\lambda_j = \sum_{n} c_{n,j} c^*_{n,j} \lambda_n $ is the effective reorganization energy,  where $c$ is the expansion coefficient, defined as $q_n=c_{n,j} q_j$.   Using Eq.~(\ref{rate}), we obtain
the VET rate from the i-th donor state to j-th acceptor state, 
\be
k = {   \la {\rm Tr} {\cal J}^2 \ra \over 2 \pi} \alpha_{A,j} \alpha_{D,i} \nu_{D,i}  (\nu_{A,j} +1)  S_{A,j} S_{D,i}
G(\omega_{A,j}-\omega_{D,i}, \lambda_{A,j} + \lambda_{D,i}) 
\ee
which will be applied to vibrational polaritons below.

Next we turn to cavity-assisted intermolecular VET under vibrational strong coupling (VSC). 
For simplicity, we consider N identical harmonic oscillators with vibrational frequency $\omega$ in a constant optical field with cavity frequency $\omega_c$.
The VSC strength is characterized by $g$ for individual oscillators and by $g_N=\sqrt{N} g$ for a collection of N oscillators.
Then, we obtain the two vibrational polariton frequencies,
\be
\omega^2_{\pm} = {1\over 2} (\omega_{e}^2  +\omega_c^2) \pm {1\over 2}
\sqrt{ (\omega_{e}^2 -\omega_c^2)^2 + g_N^2 \omega_c^4}
\label{eigen}
\ee
where the sign corresponds to the two polariton branches and 
$\omega^2_{e}=\omega^2+ g_N^2 \omega^2_c$  is the effective oscillator frequency with the account of the dipole self-energy term. \cite{cao203}
The hybrid  of the collective molecular and cavity modes is characterized by the mixing angle
\be
\tan(\theta) = {  g_N \omega_c^2 \over \omega_{e}^2 - \omega_c^2 }
\ee
where the denominator is detuning. 
The remaining $N-1$ modes are dark states with the unperturbed frequency $\omega$.   

As an example, we consider the case where both donor and acceptor are lower vibrational polariton (LP) states. 
In this case, we have the collective dipole correlation function given as
\be
C^>_{-}(\omega) & = &  {N}_{-,eff} \cos^2(\theta/2) \alpha_-^2 (\nu_- +1) G(\omega-\omega_-, \lambda_-) 
\notag
\ee
and a similar definition for $C_-^<(\omega)$.  Defined in Eq.~(\ref{eff}), $N_{eff}$ reduces to the incoherent monomer limit with $ N_{eff}=1$ 
and the fully coherent limit with $N_{eff}  = N$.   As a result, the VET rate for the two LP states becomes
\be
k = { \la  {\rm Tr} {\cal J}^2 \ra \over 2 \pi} \alpha_{A-}^2 \alpha_{D-}^2 \nu_{D-}  (\nu_{A-} +1) {N}_{A-,eff} { N}_{D-,eff} 
\cos^2(\theta_A/2) \cos^2(\theta_D/2) G(\omega_{A-}-\omega_{D-}, \lambda_{A-}+\lambda_{D-}) 
\label{cVET}
\ee
where the reorganization energy is rescaled according to $\lambda_{-} = \lambda \cos^2(\theta/2) /N$. 
Similar results can be obtained for other pairs of polaritons or molecular states.  
For molecules outside of cavity or far off-resonant with the cavity frequency,  we can set $ \cos^2(\theta/2) =1$.

\section{Cavity-assisted VET}

A recent numerical study by Li, Nitzan, and Subnotik\cite{li21} explores collective VSC effects  on vibrational energy relaxation of a small fraction of hot
molecules in the $CO_2$ solvent. As illustrated in Fig.~1(c),  the excess vibrational energy of the hot molecules is transferred transiently 
to the coherent  lower polariton state (LP) and subsequently dissipates to dark states or solvent molecules, thus 
leading to an acceleration of vibrational energy relaxation of the hot molecules.  Several intriguing  phenomena have been observed in the simulation 
and can be understood quantitatively based on the VET rate  in Eq.~(\ref{cVET}): 
\begin{enumerate}
\item
{Eq.~(\ref{cVET}) depends on the frequency detuning  via the spectral overlap (i.e., the Gaussian lineshape function) and the projection to the molecular basis (i.e., 
the prefactor), thus reproducing the bell shape in Fig.~3 of the reported simulation.
 Further, the LP wave-function yields the ratio of  $\sin^2(\theta/2)$  and $\cos^2(\theta/2)/N$ 
between the photonic energy and molecular energy,  so that the transferred energy to  
the cavity photon is higher than to molecules,  in agreement with Fig.~2(c) and Fig.~2(d) of the above reference.  
The thermalization process will eventually recover the energy equipartition for the cavity photons and molecules in the long-time limit, 
as ensured by the DB condition in Eq.~({\ref{DB}). }
\item  {The solvent concentration is linearly proportional to the Rabi frequency as in Eq.~(\ref{eigen}).   Due to static and dynamic disorders, 
quantum coherence may not extend to all solvent molecules in the LP state (i.e. donor state) 
and thus define a finite coherence scale $N_{coh}$. Thus, according to Eq.~(\ref{eff}),  the VET  rate
is linearly proportional to the solvent concentration as $N<N_{coh}$ and saturates as $N>N_{coh}$, consistent with Fig.~4(b) of the reported simulation.}
\item    {The hot molecules (i.e., donors) can be coherently coupled via the cavity field or 
the inter-molecular interaction.  Then the rate is proportional to the emission intensity, which in turn is proportional to $N_{eff}$, 
i.e.,  the number of coherently-coupled hot molecules, thus explaining the N-dependence in Fig. 5 of the referred simulation.}
\item   {Since the O-C bond is modeled as a Morse potential, its average frequency decreases with the vibrational energy. }
Then, the hot donor molecule is in better resonance with the higher vibrational excited states of the acceptor molecule, 
so the energy is transferred to the low-energy tail of the acceptor state distribution as shown in Fig.~7 of the reported simulation.}
\end{enumerate}

It is revealing to analyze the intriguing phenomena of cavity-catalyzed adiabatic 
chemical reactions in the context of enhanced vibrational energy relaxation, as suggested by the authors.\cite{li21}  
 In the current context,  the cavity-indued enhancement of the vibrational relaxation rate
of the solute molecule results in an increase in the effective
friction coefficient,  
\be
\eta_{VSC}= \eta N_{eff}, 
\ee
where $N_{eff}$ is the effective coherent number defined in Eq.~(\ref{eff}). 
According to the Kramers rate theory,\cite{hanggi90,rips90,nitzan06} 
 the $N_{eff}$-fold increase in the friction constant leads to
the enhanced reaction rate in the energy diffusion regime as $k_r \propto \eta_{VSC}$ and the suppressed
rate in the spatial diffusion regime as $k_r \propto 1/  \eta_{VSC}  $. 
The gRET theory presented here connects the reaction rate to the polariton spectral lineshape and Rabi splitting, in agreement with experiments. 
Yet, this relationship is largely temperature-independent and takes different functional forms in the under-damped and over-damped regimes.   
Thus, VSC-catalyzed reactions require further analysis, which likely involves
a quantum rate calculation.\cite{galego19,li20a,angulo20,cao203,fischer22}

Finally, we briefly comment on the validity of classical simulations of quantum dynamics.
The quantum correlation function  has a simple classical limit, which can be supplemented with
quantum correlations to ensure DB.\cite{egorov99}
 In fact, the spectrum of the harmonic potential  or the Morse potential
can be described accurately by classical dynamics,\cite{cao54}
 within the framework of  quantum-classical correspondence.\cite{cao86,cao95,gruenbaum08,loring17}
Yet, classical simulations may fail to capture more subtle quantum effects. 
An interesting example is the system-bath entanglement in the generalized FRET rate in Eq.~(\ref{FRET}). 
In a MC system,  the initial state of  the donor's emission $E(\omega)$ is an entangled
equilibrium state of system and bath and cannot be factorized. 
This non-factorized initial state brings technical difficulties in
computing the emission spectrum and FRET rate in MC systems
and has been a subject of recent studies.\cite{cao143,cao166}
Another relevant example is the reaction rate, which can be catalyzed by VSC. 
The thermal activation process involves zero-point energy shift and 
barrier tunneling, which cannot be fully captured in classical simulations and will require 
quantum mechanical analysis.\cite{galego19,li20a,angulo20,cao203,fischer22}

\section{Summary}

The generalized resonant energy transfer (gRET) theory encompasses various degrees of freedom as well as  coherently-coupled
composite states and reduces to the FRET rate in the special case of exciton transfer.  
The resulting rate expression in Eq.~(\ref{rate}) allows us to establish  the basic properties of gRET:
(i)   The rate obeys the DB condition in Eq.~(\ref{DB}), when the donor and acceptor are at the same temperature and can lead to a net flux 
when at different temperatures.
(ii)  Similar to FRET but more general, the gRET rate is proportional to the overlap between donor's emission spectrum $C^<_D$ 
and acceptor's absorption spectrum $C^>_A$. 
(iii)  The collective spectral function and rate have simple scalings with the effective coherent size defined in Eq.~(\ref{eff}),
 which is equivalent to the number of molecules for small aggregates but becomes the coherent size for large aggregates.
(iv)  The donor-acceptor coupling tensor ${\cal J}$ can be determined from the interaction potential as in Eq.~(\ref{tensor}), 
thus specifying its orientational and spatial dependences. 

The application of gRET to cavity-assisted VET provides a quantitative explanation of
 intriguing phenomena including cooperativity, resonance, and anharmonicity, as shown in a recent simulation.\c[te{li21}  
Though demonstrated in cavity-assisted intermolecular VER,  the gRET formalism can apply equally well to 
intramolecular vibrational relaxation (IVR) in cavities, as exemplified by the solvated ABA model illustrated in Fig.~1(d).\cite{wang22,cao108}
 The correlated vibrational mode coupling in polyatomic molecules may shed light on the VSC-induced phenomena
 in energy transfer and reaction kinetics and is an interesting subject for further study.

\bibliography{./bibfiles/polariton,./bibfiles/pub_cao}
\section*{Acknowledgement}
This work is supported by NSF (CHE 1800301 and CHE 1836913) and MIT Sloan Fund. 

\end{document}